# Magnon currents excited by the spin Seebeck effect in ferromagnetic EuS thin films

‡M. Xochitl Aguilar-Pujol[1], ‡Sara Catalano[1], Carmen González-Orellana[2], Witold Skowronski[1,3], Juan M. Gómez-Pérez[1], Maxim Ilyn[2], Celia Rogero[2], Marco Gobbi[1,2,4], Luis E. Hueso[1,4] and Fèlix Casanova[1,4,*]

[1]CIC NanoGUNE BRTA, 20018 Donostia-San Sebastian, Basque Country, Spain
[2]Centro de Física de Materiales CSIC-UPV/EHU, 20018 Donostia-San Sebastian, Basque Country, Spain
[3]AGH University of Science and Technology, Institute of Electronics, 30-059 Krakow, Poland
[4]IKERBASQUE, Basque Foundation for Science, 48009 Bilbao, Basque Country, Spain
*E-mail: f.casanova@nanogune.eu
‡These authors contributed equally

## ABSTRACT

A magnetic insulator is an ideal platform to propagate spin information by exploiting magnon currents. However, until now, most studies have focused on $Y_3Fe_5O_{12}$ (YIG) and a few other ferri- and antiferromagnetic insulators, but not on pure ferromagnets. In this study, we demonstrate that magnon currents can propagate in ferromagnetic insulating thin films of EuS. By performing both local and non-local transport measurements in 18-nm-thick films of EuS using Pt electrodes, we detect magnon currents arising from thermal generation by the spin Seebeck effect. By comparing the dependence of the local and non-local signals with the temperature (< 30 K) and magnetic field (< 9 T), we confirm the magnon transport origin of the non-local signal. Finally, we extract the magnon diffusion length in the EuS film (~140 nm), a short value in good correspondence with the large Gilbert damping measured in the same film.

## I. INTRODUCTION

Magnons, the collective bosonic excitations in magnetically ordered systems, can propagate and transport spin angular momentum even in insulators [1,2]. However, to pave the way for pure spin-based information, spin signals in magnetic insulators must integrate with conventional electronics [1]. Therefore, the ideal platform for these devices comprises the interface between a metal and a magnetic insulator. This arrangement enables the transport of spin angular momentum between the electrons of the metal and the magnons in the magnetic insulator via the interfacial exchange interaction, which is quantified by the spin-mixing conductance. Incoherent magnon currents can be excited both electrically, by means of the spin Hall effect (SHE), or thermally, due to the spin Seebeck effect (SSE) [1–3]. In order to detect these magnon currents, the inverse spin Hall effect (ISHE) can be used, since it transforms back a spin current into a charge current, enabling all-electrical access to spin current physics [2,3].

So far, magnon transport in magnetic insulators has been studied mainly through ferrimagnetic garnets with the prototypical example being $Y_3Fe_5O_{12}$ (YIG), whose exceptionally small Gilbert damping results in a magnon diffusion length of several microns [4–6]. Magnon transport has also been reported in some antiferromagnetic insulators, showing characteristic magnon diffusion lengths of few hundreds of nm [7,8]. Indeed, there is increasing interest in extending the knowledge of magnon transport to other magnetic compounds. For example, recent studies have focused on magnon excitations in Van der Waals magnetic insulators [9–12]. In this context, spin Hall magnetoresistance (SMR) measurements on Eu-based insulating ferromagnets have recently demonstrated intriguing spin-transport properties at the interface with heavy metal films, suggesting that they could also be employed as carriers of magnon spin currents [13–15].

Europium sulphide (EuS), one of the few examples of isotropic Heisenberg ferromagnetic insulator (FI) [16,17], can be grown as thin films exhibiting the ferromagnetic ground state below the Curie temperature $T_C \approx 18$ K, which can be tuned by chemical doping or strain [18]. Below $T_C$, it behaves as a soft ferromagnet with extremely small coercive fields, similar to YIG. EuS films have been used to introduce strong magnetic exchange fields within interfacial layers such as metals, superconductors, and topological insulators, and manipulate their electronic phases by the magnetic proximity effect [15,19–30]. Specifically, EuS/Pt interfaces have recently been studied by means of SMR measurements, revealing a strong exchange field into the heavy metal layer, even for polycrystalline EuS films [15].

In this paper, we demonstrate the propagation of magnon spin currents in a ferromagnetic insulating thin film of EuS. We use Pt nanostructures to generate and detect magnon currents through evaporated polycrystalline EuS films. Below $T_C$ of EuS, we show that magnon currents generated by the SSE at the Pt/EuS interface can propagate through the EuS films. We study such an effect by electrically detecting the magnon currents at the Pt/EuS interface considering both the local and the non-local configuration. We study the temperature, magnetic field, and length dependence of the signal, which indicate that thermally induced magnon currents propagate in the diffusive transport regime in our samples. We extract a thermal magnon diffusion length $\ell_m^{th}$ of ~140 nm, much smaller than the one observed in YIG, suggesting that magnons are strongly damped in the studied EuS films, in good correspondence with the measured Gilbert damping $\alpha_G$ (~0.04). Our work further expands the present knowledge of magnon transport to a broader class of materials.

## II. EXPERIMENTAL DETAILS

Pt/EuS heterostructures have been fabricated on top of insulating Pyrex substrates, following the procedure presented in our previous work [15]. First, 5 nm of Pt were deposited with DC magnetron sputtering on top of the Pyrex substrate [31]. Subsequently, the Pt/EuS magnon spin transport (MST) devices were defined by e-beam lithography. Each MST device consists of two or three Pt strips (width of 300 nm and length of 70 μm) separated by different distances $d$ (0.8 μm < $d$ < 2 μm). Afterwards, 18 nm of EuS were evaporated *ex situ* on top of the Pt contacts, with the same deposition method reported by Gomez-Perez et al. [15]. Since the top surface of the EuS film oxidizes when it is exposed to air, the final film corresponds to around 14 nm of insulating EuS capped by ~4 nm of EuO$_x$, which is also insulating [32]. An optical image of a representative device is shown in Fig. 1(d).

Transport measurements were carried out using a Quantum Design Physical Property Measurement System (PPMS) covering the temperature range 2 K < $T$ < 300 K. We applied magnetic fields $H$ up to $\mu_0 H$ = 9 T and sample was rotated in the xy plane [$\alpha$ plane, inset of Fig. 1(a)]. We applied a DC current $I$ in the range 4 µA < $I$ < 100 µA with a Keithley 6221 current source meter and measured the voltage with a Keithley 2182 nanovoltmeter. As sketched in Fig. 1(d), $I$ is applied through one Pt strip, and we studied the voltage response in two configurations. We either detect the voltage along the same strip, which we refer to as local voltage $V_{LOC}$, or we detect it in a second Pt strip separated by a distance $d$, that we denote as a non-local voltage $V_{NL}$. We analysed the first and second harmonic components of the detected voltages $V_{LOC}$ and $V_{NL}$ by applying the DC current reversal technique, or delta mode [33,34]. We extracted the first harmonic or electrical response component from $V^e_{LOC,NL} = (V_{LOC,NL}(I+) - V_{LOC,NL}(I-))/2$. The second harmonic component, or thermal voltage response, is provided by $V^{th}_{LOC,NL} = (V_{LOC,NL}(I+) + V_{LOC,NL}(I-))/2$ [4,33–35].

The magnetic properties of the EuS film have been measured with the vibrating sample magnetometer (VSM) and ferromagnetic resonance (FMR) options of the Quantum Design PPMS.

## III.  RESULTS AND DISCUSSION

We studied the generation and transport of spin angular momentum in EuS films by incoherent magnon currents, which can be driven by non-equilibrium magnon density and temperature gradients [36]. In fact, a magnon spin current $j_m$ can propagate in a magnetic medium according to the equation:

$$\frac{2e}{\hbar} j_m = (-\sigma_m \nabla \mu_m + \frac{L}{T} \nabla T_m)$$

where $\sigma_m$ is the magnon spin conductivity, $\mu_m$ is the non-equilibrium magnon chemical potential, $L$ is the spin Seebeck coefficient of the medium, $T$ is the average equilibrium temperature of the magnon bath and $\nabla T_m$ the temperature gradient applied to the system [37].

In our experiment, we adopted the same measurement configuration used by Cornelissen et al. [4]. The magnon spin transport (MST) devices consist of a EuS thin film deposited on top of the Pt strips. As sketched in Fig. 1(d), a charge current $I$ is applied through a metallic Pt strip (injector) and the voltage response is measured along the same strip (local voltage, $V_{LOC}$) or at a different strip (non-local voltage, $V_{NL}$). We recorded both the first ($V^e_{LOC}$, $V^e_{NL}$) and second harmonic ($V^{th}_{LOC}$, $V^{th}_{NL}$) response with the DC current reversal technique [33,34], as described in Sec. II.

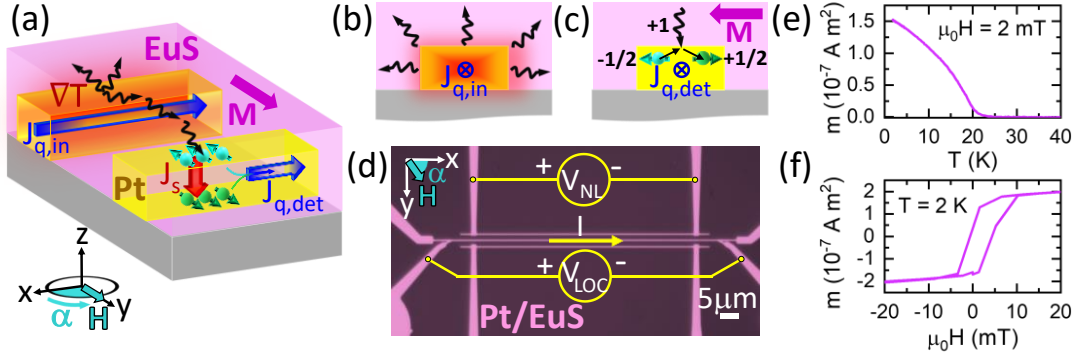

FIG. 1. (a) Schematic representation of the physical processes that occur in the Pyrex/Pt/EuS system. Two Pt strips are embedded in the magnetic insulator (EuS) with magnetization $M$ following the in-plane magnetic field $H$ applied, which can rotate an angle $\alpha$ in the sample plane. When a charge current density $j_{q,in}$ is applied through the left Pt strip, a radial thermal gradient $\nabla T$ appears due to Joule heating, generating non-equilibrium magnons (black arrows) that diffuse away. This leads to a magnon accumulation at the second Pt strip that interact with the electron-spins of the heavy metal creating a spin accumulation that induces a spin current density $j_s$. Due to the ISHE, $j_s$ is transformed into a charge current density $j_{q,det}$ that can be electrically detected. (b) The applied $j_{q,in}$ generates a thermal gradient due to ohmic losses that gives rise to a magnon flow away from the injector. (c) Spin-flip scattering process leading to a transfer of spin angular momentum between the Pt electrons and the EuS magnons at the Pt/EuS interface. (d) Optical microscope image of a device, showing the measurement configuration where $I$ denotes the charge current applied, and $V_{LOC}$, $V_{NL}$ the local and non-local voltages measured, respectively. (e) Temperature dependence of the magnetic moment measured at a magnetic field of 2 mT in the Pyrex/Pt/EuS sample. (f) Magnetic hysteresis loop measured in the Pyrex/Pt/EuS sample at 2 K, with a step size of 2.5 mT.

The measurements principle is illustrated in Fig. 1(a). The applied $I$ along the x-axis of the Pt wire corresponds to a charge current density $j_{q,in}$ that generates a spin accumulation polarized along the y-axis at the Pt/EuS interface thanks to the SHE. The effects of the interaction between the spin current flowing in the Pt strip in z-axis and the interfacial magnetic layer can be read out in the first order, or *electrical*, response of the system as a voltage $V^e_{LOC,NL}$. By measuring $V^e_{LOC}$ we study the spin Hall magnetoresistance (SMR) effect in our samples, that is the modulation of the Pt resistance due to the torque exerted by the magnetization ($M$) of the EuS on the spin current flowing through Pt [38,39]. Moreover, the spin current can also generate magnons at the Pt/EuS interface by spin-flip scattering processes, as sketched in Fig. 1(c). Furthermore, when the magnons diffuse away and reach a second Pt strip, they can transfer spin angular momentum to the Pt electrons due to the same spin-flip scattering process [Fig. 1(c)]. This produces a spin current through the Pt/EuS interface and, consequently, a voltage in the Pt wire, due to the ISHE. Thus, the electrically injected magnon currents are detected as $V^e_{NL}$ in the second Pt wire.

Most important for our work, the applied $j_{q,in}$ also generates Joule heating at the Pt/EuS interface, so that a temperature gradient proportional to the square of the current ($I^2$) is introduced in the system [Fig. 1(b)]. The second order or *thermal* response of the system to such a gradient is detected as $V_{LOC,NL}^{th}$. When the thermal gradient appears, the magnon population is driven out of equilibrium and a magnon current can flow between the hot and the cold side of the system [black arrows in Figs. 1(a), 1(b), and 1(c)], resulting in the SSE. At the interface with the Pt wire, the magnon spin angular momentum is transferred to the Pt electrons through spin-flip scattering processes [Fig. 1(c)], yielding a thermal voltage response $V_{LOC,NL}^{th}$. We note here that the thermal voltage due to the SSE can be detected both locally ($V_{LOC}^{th}$) and non-locally ($V_{NL}^{th}$), as thermally induced magnon currents [black arrows in figures 1(a), 1(b), and 1(c)] diffuse through the EuS film [1,40,41].

The geometry of the experiment allows the SMR and magnon-induced voltages to be probed through the Pt wires. We take into account that both phenomena depend on the orientation between the spin polarization ***s*** of the electrons in the Pt (fixed along y axis), and $M$ of EuS. For that reason, we saturate $M$ in the plane of the film (xy-plane) with an external magnetic field $H$ [inset of figure 1(a)], and we rotate it. Thus, since the generated spin current is absorbed by $M$ as a spin-transfer torque if $M \perp s$, SMR results in a $\cos^2 \alpha$ angular dependence when $M$ of the EuS film is rotated in plane [Fig. 1(a)] [42]. However, spin-flip scattering processes depends on the scalar product between $M$ and ***s***, causing the electrically injected and detected magnon currents to exhibit a $\sin^2 \alpha$ dependence [4]. Finally, since spin-flip only occurs at the detector for the thermal injection, it results in a $\sin \alpha$ dependence [4].

The magnetic properties of the studied EuS film are exemplified in Figs. 1(e) and 1(f), which present the total magnetic moment ($m$) of the sample as a function of temperature and applied magnetic field, respectively. As shown in Fig. 1(e), EuS exhibits a clear ferromagnetic behavior with a $T_C \approx 19\ K$, in agreement with previous reports [15,18]. The hysteresis loop at 2 K in Fig. 1(f), with a coercive field around 3 mT, confirms the ferromagnetic behavior of the EuS film (see the Supplemental Material [43], Sec. S5 for more details) [15].

### A. Angle dependence of the electrical and thermal response.

Figure 2(a) presents the angular-dependent magnetoresistance (ADMR) measured in MST1 as a function of the in-plane angle $\alpha$, at $T = 2$ K, below $T_C$ of EuS. We saturate $M$ in plane by applying a small external field ($\mu_0 H = 0.1$ T). The data are extracted from the electrical local response $V_{LOC}^{e}$ in order to provide the SMR signal $\Delta\rho_L/\rho = [R_L(\alpha) - R_L(90º)]/R_L(90º)$, where $R_L = V_{LOC}^{e}/I$ is the longitudinal resistance. We observe a clear $\cos^2(\alpha)$ modulation of the Pt resistance, as expected for the SMR effect [38,39,47]. The signal amplitude [double arrow in fig 2(a)] is of the order of $10^{-4}$, consistent with previous results in EuS/Pt interfaces and comparable to the magnitude of the SMR measured in Pt/YIG interfaces [15,47,48]. Field-dependent magnetoresistance (FDMR) measurements confirm the typical $M$ behavior of EuS, displaying the SMR gap and magnetoresistance peaks in correspondence with the expected magnetization reversal of EuS (see the Supplemental Material [43], Sec. S1). The large SMR amplitude and clear correlation with $M$ of EuS indicate an efficient spin transfer at the EuS/Pt interface, in other words, a favorable spin-mixing conductance [14,15,49]. We note that, from the SMR measurements, we can infer the spin transfer efficiency of each device, which we use to normalize the data to compare the response of different devices, as described in the Supplemental Material [43], Sec. S2. Hereafter, we label the normalized measured voltage $V_{LOC}^*$ and $V_{NL}^*$.

Next, we study the magnon currents. For these measurements, a current $I \leq 20$ μA is applied through an injector strip as sketched in the right panel of Figs. 2(b) and 2(c). Note that the injected current is chosen to be small in order to keep the local temperature below the Curie point of the EuS film, as verified by measuring the four-point resistance of the Pt injector. A detailed calibration of the injector temperature with respect to the applied current is provided in the Supplemental Material [43], Sec. S2.

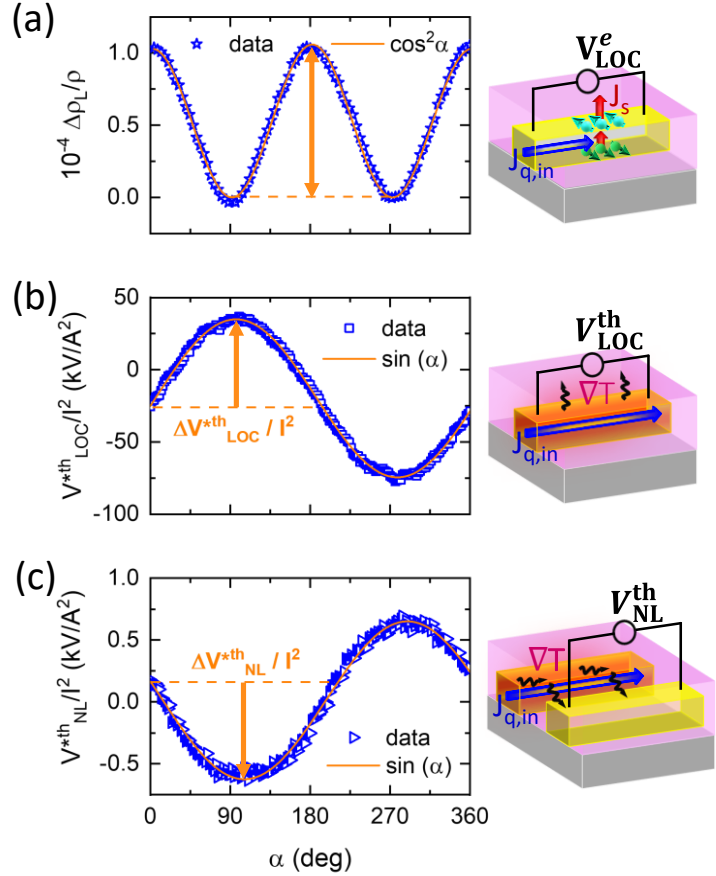

FIG. 2. Data corresponds to device MST1. Representative (a) ADMR of the SMR at a fixed magnetic field of $\mu_0 H = 0.1$ T. (b) Local SSE and (c) non-local SSE at 0.8 μm, both at $\mu_0 H = 0.3$ T, where the voltage is normalized to the square of the applied current. The measurement temperature is 2 K and the current applied is 20 μA. The arrows indicate the amplitude of the signal. A schematic illustration of the effect measured is included on the right side of the corresponding panel.

First, we measure the thermal local amplitude $V_{LOC}^{th}$ (which we normalize to $I^2$) as a function of $\alpha$ at $T = 2$ K and $\mu_0 H = 0.3$ T. $V_{LOC}^{th}/I^2$ shows a clear $\sin \alpha$ modulation, as presented in Fig. 2(b) for device MST1. The $\sin \alpha$ angular dependence is consistent with the symmetry expected for the SSE [4]. A constant angle-independent voltage offset is also present due to other thermoelectric effects [4,6,8,50]. Secondly, we study the propagation of magnon currents through the non-local voltage. For the electrically injected magnon currents, that follows a $\sin^2\alpha$ dependence, we found no signal in $V_{NL}^e$ (Supplemental Material [43], Sec. S3) at any of the measured distances 0.8 μm < $d$ < 2 μm. In contrast, $V_{NL}^{th}/I^2$ signal is similar to the local $V_{LOC}^{th}/I^2$ with the same clear $\sin\alpha$ modulation expected for the SSE, but with opposite sign, as shown in Fig. 2(c) for MST1 with $d$ = 0.8 μm. The offset signal due to other thermoelectric effects is also present for the non-local case. The $\sin\alpha$ angular dependence, verified at all the measured distances for $V_{NL}^{th}/I^2$, is consistent with the symmetry expected for SSE-induced magnon currents [4,34,35]. However, the inverted sign suggests that magnon currents at the injector/detector strips are flowing in opposite directions [Figs. 2(b) and 2(c)]. In fact, such a sign change is expected to occur between the SSE induced magnon currents measured locally and non-locally, due to the redistribution of the magnon population induced by Joule heating [51–53]. As Joule heating deplete the magnon distribution at the injector site, the magnon currents are expected to flow towards (away from) the detector (the injector) as also illustrated by the direction of the black arrows in Figs. 1(b) and 1(c) [37,51,52]. For all the devices measured, we observed a positive local $V_{LOC}^{th}/I^2$ amplitude and a negative non-local $V_{NL}^{th}/I^2$ amplitude.

All things considered, the Pt/EuS devices reveal a magnon-induced response to Joule heating, consistent with the symmetry of the SSE. The absence of an analogue signal in the non-local electrical response $V_{NL}^e$, in contrast, suggests that the magnon population redistribution induced by the spin-flip scattering process at the injector Pt/EuS interface is too small to produce a measurable signal at the detector. We note here that $V_{NL}^e/I$ has been reported to vanish as the temperature is lowered below 50 K in Pt/YIG interfaces by different groups [35,54,55]. Thus, the absence of a signal $V_{NL}^e/I^2$ for our samples is expected in the measured temperature range.

### B. Temperature dependence

Subsequently, we analyze the temperature dependence of the SMR, local SSE and non-local SSE amplitudes [defined in Fig. (2)], which are presented in Fig. 3 for devices MST1 and MST3. In all cases, the amplitude of the signal is maximum for the lowest temperature measured (2 K) and decreases with increasing temperature. However, we observe a clear difference between the SMR curve and the thermal amplitudes $V_{LOC}^{*th}/I^2$ and $V_{NL}^{*th}/I^2$, as can be observed in Fig. 3(d). The temperature dependence of the SMR amplitudes [Fig. 3(a)] follows the expected trend vanishing as the temperature is raised above the Curie point of the EuS film [see Fig. 1(e) and the Supplemental Material [43], Fig. S7]. In fact, we note here that we can measure an SMR signal even above the $T_C$ of the EuS films. The presence of such a finite SMR response above $T_C$ is a consequence of the sensitivity of the SMR to the magnetic correlations [56], which are present even above $T_C$ in our films, according to our VSM characterization (Supplemental Material [43], Sec. S5). The observed temperature dependence of SMR in our EuS thin films can be fully explained in terms of the microscopic theory developed by Zhang et al. [15,57]. Instead, the temperature dependence of $V_{LOC}^{*th}/I^2$ [Fig. 3(b)] and $V_{NL}^{*th}/I^2$ [Fig. 3(c)] shows similar trend with a substantially different decrease from the SMR curve as $T_C$ is approached [Fig. 3(d)]. Such a temperature dependence may seem surprising for SSE induced thermal voltages, for which the SSE theory predicts a linear dependence with the system magnetization [58]. However, a similar behavior has been reported in other experimental studies of the SSE at low temperatures [35,54] and can be qualitatively understood with the following considerations. First, the Joule heating induced temperature gradient is very likely temperature dependent, as the thermal conductivities of Pt, EuS and the Pyrex substrate may strongly change and at different rates as the temperature is lowered. Second, the Gilbert damping of the EuS film also varies at such temperatures (see the Supplemental Material, Fig. S8) due to the very low $T_C$ of the films, which consequently affects the propagation of magnon currents. Elaborating a model that captures such temperature dependent effects on the measured thermal voltages is a challenging task, which goes beyond the scope of this work. Instead, we remark that all the measured signals disappear above the $T_C$ of the EuS film, confirming the magnetic origin of the studied voltages.

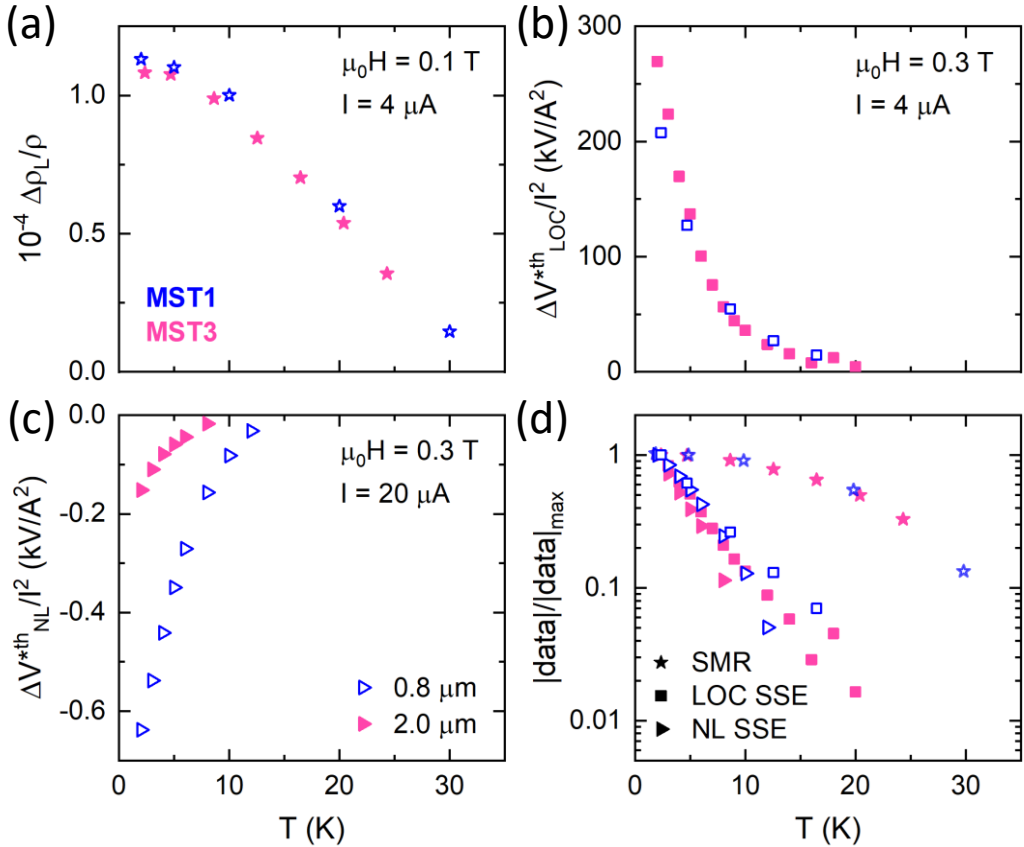

FIG. 3. Data corresponds to devices MST1 (blue open symbols) and MST3 (pink solid symbols). (a) SMR amplitudes [as defined in Fig. 2(a)] at $\mu_0 H = 0.1$ T and $I = 4$ µA, (b) local SSE amplitudes [as defined in Fig. 2(b)] at $\mu_0 H = 0.3$ T and $I = 4$ µA, and (c) non-local SSE amplitudes [as defined in Fig. 2(c)], at $\mu_0 H = 0.3$ T and $I = 20$ µA for two different distances, as a function of temperature. (d) Comparison of SMR, local SSE and non-local SSE amplitudes normalized to their maximum values.

## C. Magnetic field dependence

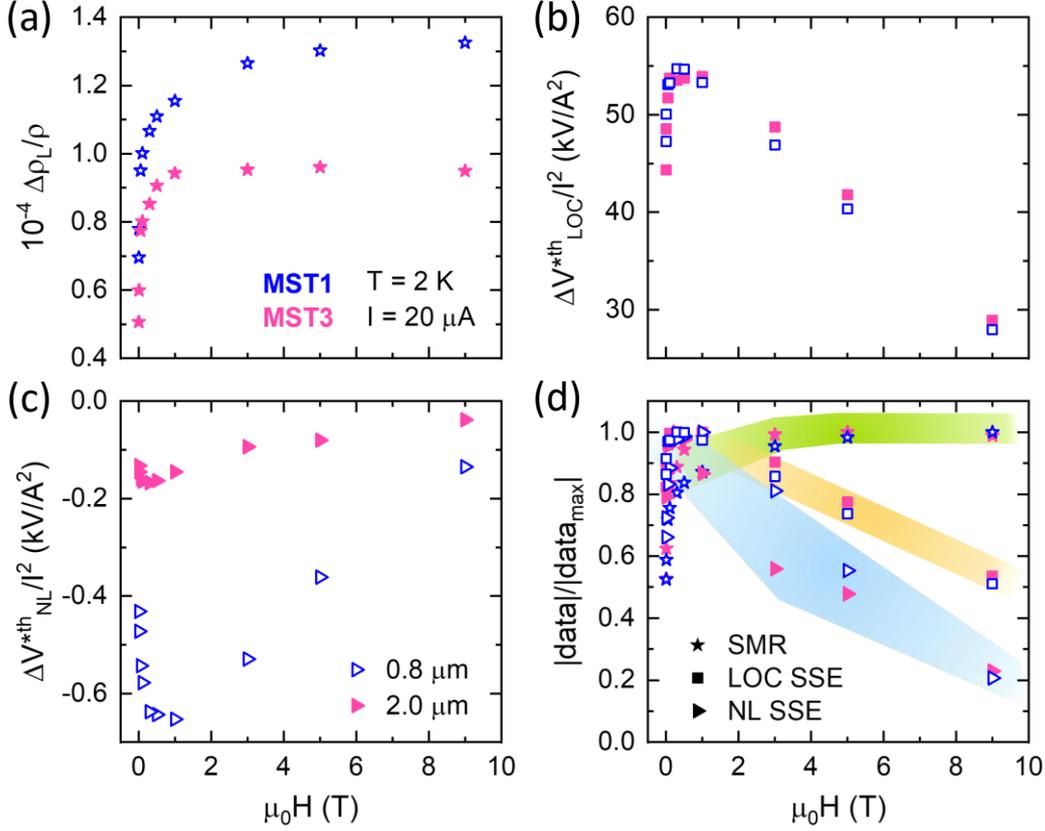

FIG. 4. Data corresponds to devices MST1 (blue open symbols) and MST3 (pink solid symbols). (a) SMR, (b) local SSE, and (c) non-local SSE amplitudes (as defined in Fig. 2) as a function of the magnetic field at $T = 2$ K and $I = 20$ μA. (d) Comparison of SMR, local SSE, and non-local SSE amplitudes normalized to their maximum values. Colored shadows are a guide to the eye, being green for the SMR amplitudes, orange for the local amplitudes and blue for the non-local amplitudes.

We also examine the magnetic field dependence of the SMR, local SSE and non-local SSE amplitudes, as illustrated in Fig. 4. Initially, both the SMR and SSE signals increase as $M$ of the system develops, since both effects are related to the interaction of the spins of the electrons in Pt with the magnetic moments in the magnetic layer; but as $M$ continues to grow their behavior diverges. The SMR response tends to saturate with the magnetic field [Fig. 4(a)], in correspondence with the saturation of $M$ in the EuS films. Note here that the saturation field of EuS is much higher than the observed coercive field [Fig. 1(f)], as already observed in films of EuS deposited at room temperature [20,59]. In contrast to the SMR, both SSE curves [Figs. 4(b) and 4(c)] reach a maximum at $\mu_0 H \approx 0.5$ T, followed by a gradually reduction of the SSE amplitude for higher magnetic fields. Such a decay is characteristic of SSE induced voltages, for which the opening of the Zeeman gap affects the magnon population. According to the SSE theory, when the Zeeman energy $g\mu_B H$ is larger than the thermal energy $k_B T$, magnons cannot be thermally excited, leading to the suppression of the local SSE, where $g$, $\mu_B$ and $k_B$ are the $g$-factor, Bohr magneton and Boltzmann constant, respectively [60–62]. More interestingly, we note that the local SSE signal suppression at the maximum applied field ($\mu_0 H = 9$ T) is almost 50% of the maximum signal whereas non-local SSE signal suppression reaches 80%, as is shown in Fig. 4(d). Such a difference can be accounted for by the distinct way magnon currents reach the EuS/Pt interface in the local and non-local case. In fact, the magnon induced non-local voltage should decay exponentially on the scale of the magnon diffusion length, in contrast to the local voltage case. Therefore, as the magnon diffusion length is also suppressed as the magnetic field is increased, a stronger suppression of the signal should be expected for the non-local case [63].

### D. Magnon diffusion length

Finally, we study the dependence of the non-local SSE amplitude on the injector-detector distance *d*, to unravel the mechanism by which the magnon currents propagate through the EuS medium. As shown in Fig. 5, the data can be fitted with an exponential decay law, characteristic of thermally generated magnons in the relaxation regime [4,64]. Moreover, we note that the data could not be fitted by power laws, indicating that the signal is not driven by the radial decay of a thermal gradient through the sample, but by the redistribution of the magnon population (see Supplemental Material [43], Sec. S4). From the exponential fit we extract a magnon diffusion length $\ell_m^{th} = 140 \pm 30$ nm at 2 K and 0.3 T. This value is of the same order as the one reported by Gao *et al.* in the ferrimagnet TmIG, where a 15-nm-thick film results in $\ell_m^{th} = 300$ nm at 0.5 T and room temperature [65]. In comparison to the best YIG samples, where $\ell_m^{th} \sim 7$ µm at low temperatures in a 210-nm-thick film and the electrical magnon diffusion length $\ell_m^e \sim 3$ µm in 10–15-nm-thick films [6,54], the EuS value is more than one order of magnitude smaller. However, we note here that several studies in YIG films reported $\ell_m^{th}$ values at room temperature comparable with the $\ell_m^{th}$ value that we observe in EuS [66–69]. At this point, it is important to consider the Gilbert damping ($\alpha_G$) since it is linked to the magnon diffusion length. Reducing $\alpha_G$ implies increasing the magnon spin-relaxation time and, therefore, the magnon diffusion length [2,37]. To verify this relationship, we performed FMR measurements in the very same sample where the measured devices are located. Our results in the EuS thin film yields $\alpha_G^{EuS} = 4 \times 10^{-2}$ at 2 K (see the Supplemental Material [43], Sec. S5), which is two orders of magnitude higher than the case of YIG thin films ($\alpha_G^{YIG} \sim 10^{-4}$) [4,6], in agreement with the difference observed in the magnon diffusion length.

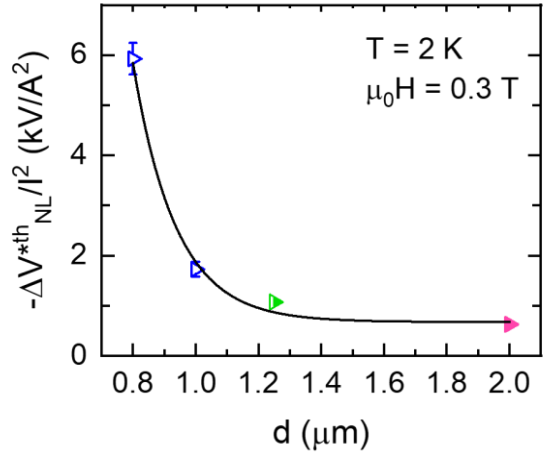

FIG. 5. Amplitude of the non-local SSE as a function of injector-detector separation distance measured at $T$ = 2 K, $\mu_0 H$ = 0.3 T and $I$ = 4 µA. Black solid line is a fit to an exponential decay.

## IV. CONCLUSIONS

To summarize, we demonstrated the transport of incoherent magnon currents in polycrystalline films of EuS. Below $T_C$, we observe a thermally induced response in both local and non-local transport measurements. Considering the angular, temperature and magnetic field dependence of such signals, we ascribe the measured thermal voltages to the SSE. By studying the length dependence of the non-local transport signal, we extract a magnon diffusion length of $\ell_m^{th} = 140 \pm 30$ nm at 2 K and 0.3 T. This value, short compared to other studied materials such as YIG, correlates well with an enhanced Gilbert damping caused by the polycrystalline structure of the studied EuS films. While we expect that a significantly smaller Gilbert damping in epitaxial EuS films could lead to an improvement of the magnon diffusion length, we highlight the observation of magnon currents propagating even through a polycrystalline ferromagnetic insulator film, as compared to the case of YIG films. Despite the relatively short $\ell_m^{th}$ observed, our work shows that the transport of spin currents by incoherent magnons through EuS films should be taken into account when studying EuS-based heterostructures. As for perspectives, our results evidence the opportunity of studying the interaction between magnon currents and superconductivity in systems comprising, for example, superconductor/EuS interfaces, following recent experimental and theoretical works [70,71]. Moreover, as sizeable thermoelectric effects may occur in spin-split superconductors, such as Al/EuS bilayers, the role of the observed SSE could be further explored in such a context [72–75].

## ACKNOWLEDGEMENTS

We would like to thank Mihail Ipatov for technical expertise in the FMR measurements. We acknowledge funding by the Spanish MICINN (project No. PID2021-122511OB-I00 and Maria de Maeztu Units of Excellence Programme No. CEX2020-001038-M) and by the European Union H2020 Programme (Projects No. 965046-INTERFAST and 964396-SINFONIA). M.X.A.-P. thanks the Spanish MICINN for a Ph.D. fellowship (grant No. PRE-2019-089833). S.C. acknowledges support from the European Commission for a Marie Sklodowska-Curie individual fellowship (Grant No. 796817-ARTEMIS). W.S. acknowledges financial support from Polish National Agency for Academic Exchange (PPN/BEK/2020/1/00118/DEC/1).

# SUPPLEMENTAL MATERIAL

**S1. Longitudinal magnetoresistance measurements.**

Figure S1(a) shows a characteristic longitudinal field-dependent magnetoresistance (FDMR) measurement performed in device MST2 at $T = 2$ K with the magnetic field applied in x and y direction. A clear gap appears between $H_x$ and $H_y$ curves with a peak (in $H_y$) or dip (in $H_x$) around zero field which correspond to the reversal magnetization of EuS and follows the SMR behavior for in-plane magnetic fields [1]. Figure S1(b) shows the angular-dependent magnetoresistance (ADMR) measurement for the same device at $T = 2$ K and $\mu_0 H = 0.1$ T, where the amplitude of the signal corresponds to the magnitude of the FDMR gap at the same magnetic field.

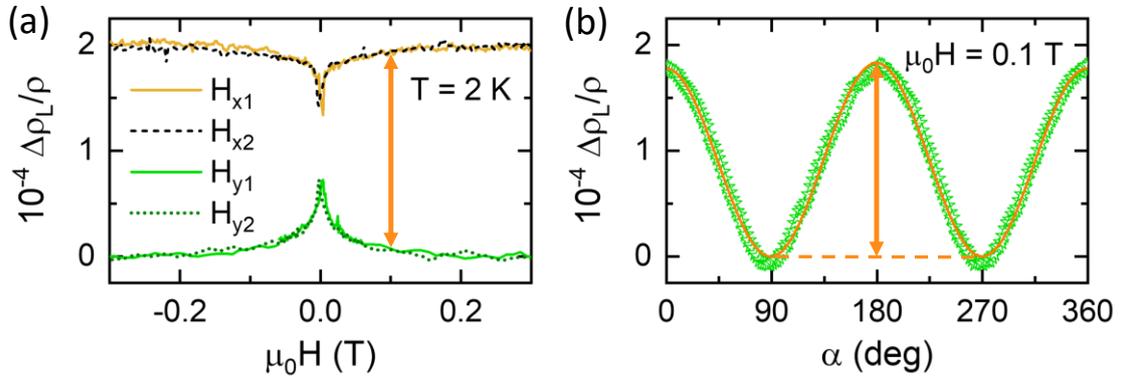

**FIG. S6**. Longitudinal measurements of the (a) FDMR at $T = 2$ K and (b) ADMR in $\alpha$ plane at $T = 2$ K and $\mu_0 H = 0.1$ T in device MST2. The orange line is a $cos^2 \alpha$ fitting. The double arrows in (a) and (b) mark the same SMR amplitude $\Delta \rho_L / \rho$.

**S2. Current-dependent measurements**

Figure S2(a) shows the ADMR measurement performed for each device (MST1, MST2 and MST3) at $I = 4$ $\mu$A, $\mu_0 H = 0.1$ T and $T = 2$ K. The different amplitude among the devices implies a different spin transfer at the EuS/Pt interface (i.e., a different effective spin-mixing conductance). Therefore, in order to normalize the measured voltage $V_{LOC,NL}$ at a given current, we use the following formula:

$$(V^*_{LOC,NL})_{MSTx} = (V_{LOC,NL})_{MSTx} \cdot \frac{(\Delta \rho_L / \rho)_{MST2}}{(\Delta \rho_L / \rho)_{MSTx}} \tag{S1}$$

where $\Delta \rho_L / \rho$ is the amplitude of the SMR signal and $MSTx$ a device. As shown in Fig. S2(b), the amplitude of the SMR signal decreases as the current is increased.

Since the injected current heats up the Pt injector and the Curie temperature ($T_C$) of the EuS is low, we calibrate the temperature of the injector to stay below $T_C$. To do so, first we have measured the temperature dependence of the Pt at the lowest current applied (4 $\mu$A), a representative curve for device MST2 is shown in figure S3(a). Then, from the minimum of the SMR curve at higher currents, up to 100 $\mu A$, we estimate the effective temperature of the system [see figures S3(b) and S3(c)]. This way we make sure that we are not above the Curie temperature of the EuS, neither increasing too much the temperature of the system. As shown in figure S3(c), a linear dependence between the applied current and the effective temperature is found. We ascribe the linear relationship not only to Joule heating, which is proportional to $I^2$, but also to the heat removed from the sample surface by the cooling capacity of the He-vapor based cryostat, which is also strongly temperature dependent below 20 K.

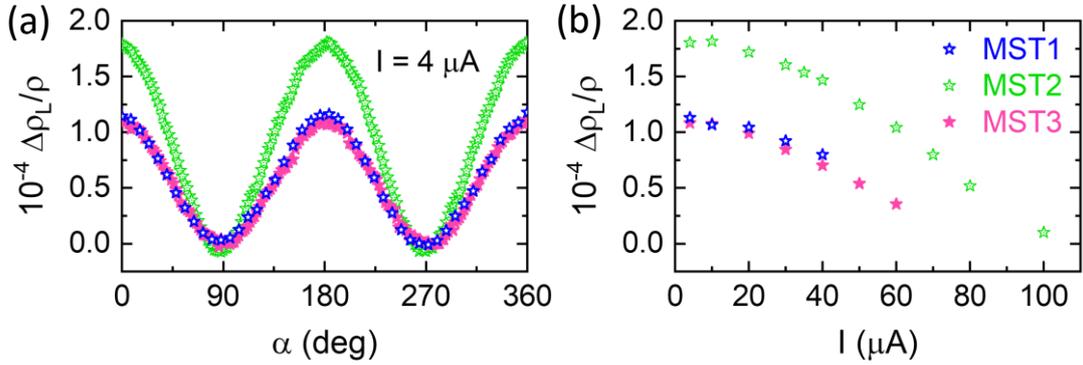

**FIG. S7**. Data corresponds to devices MST1 (blue open stars), MST2 (green open stars) and MST3 (pink solid stars). (a) ADMR at $T = 2$ K, $\mu_0 H = 0.1$ T and the smallest current applied (4 $\mu A$). (b) Current dependence of the ADMR amplitude at $T = 2$ K and $\mu_0 H = 0.1$ T.

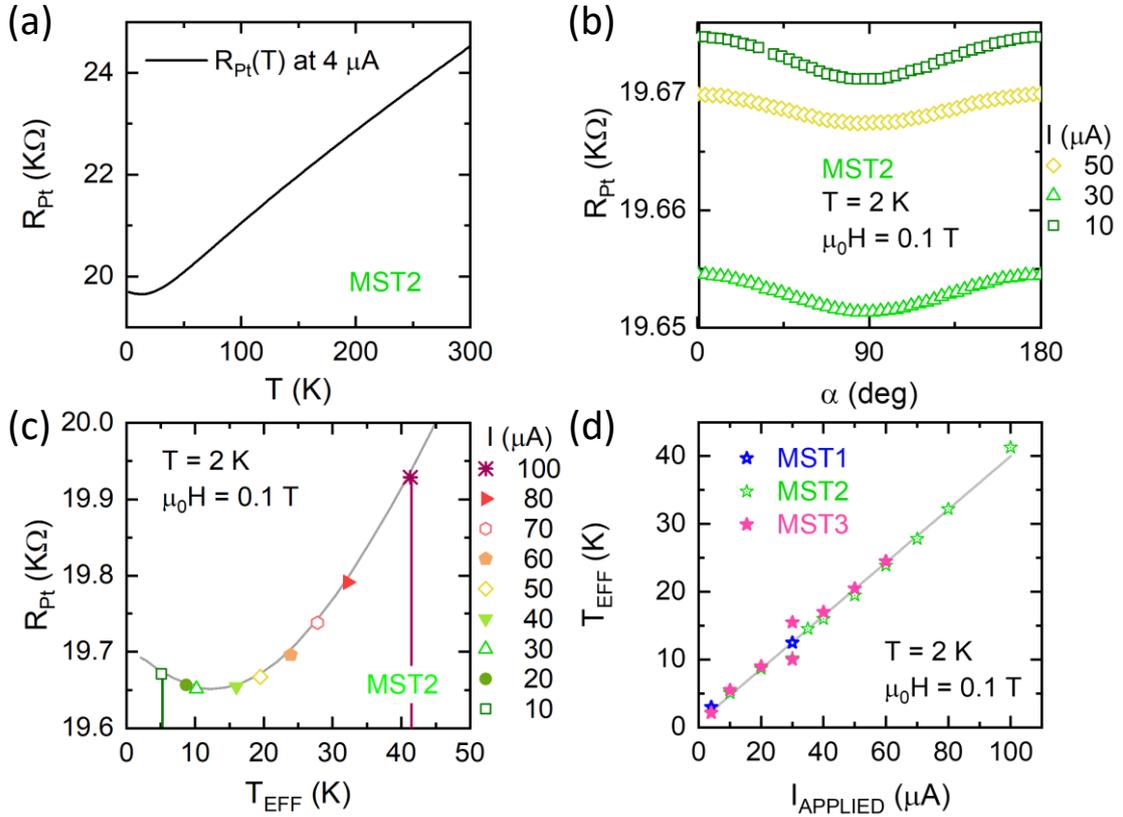

**FIG. S8**. (a) Resistance of Pt as a function of temperature at $I = 4$ $\mu$A measured in the longitudinal configuration. (b) ADMR measurements in $\alpha$ plane for $I = 10$, 30 and 50 $\mu$A. The minimum of the signal (at $\alpha = 90°$) changes with the applied current. The SMR amplitude decreases with increasing applied current, which can be better seen in Fig. S2(b). (c) Correspondence between the resistance of Pt as a function of temperature at 4 $\mu$A (grey line) and the minimum of the SMR curves measured at different currents (colored symbols). Data in panels (a), (b), and (c) correspond to device MST2. (d) Relation between the current applied and the effective temperature [extracted from panel (c) for the exemplary case of MST2]. The same linear relation (grey solid line) is obtained for the three studied devices (MST1, MST2, and MST3).

Moreover, due to the relatively low $T_C$ of the EuS films, the second harmonic voltage is particularly affected by Joule heating. As can be seen in Fig. S4, the voltage versus $I^2$ characteristics reveals two competing regimes. At low currents, we see an increase of voltage with the squared current, but as we continue increasing the current, the signal reaches a maximum and starts decreasing. This is most likely due to overheating of the EuS layer, and in consequence, reducing the magnetization of the film. For that reason, we keep at $I \leq 20$ µA for the local and non-local measurements.

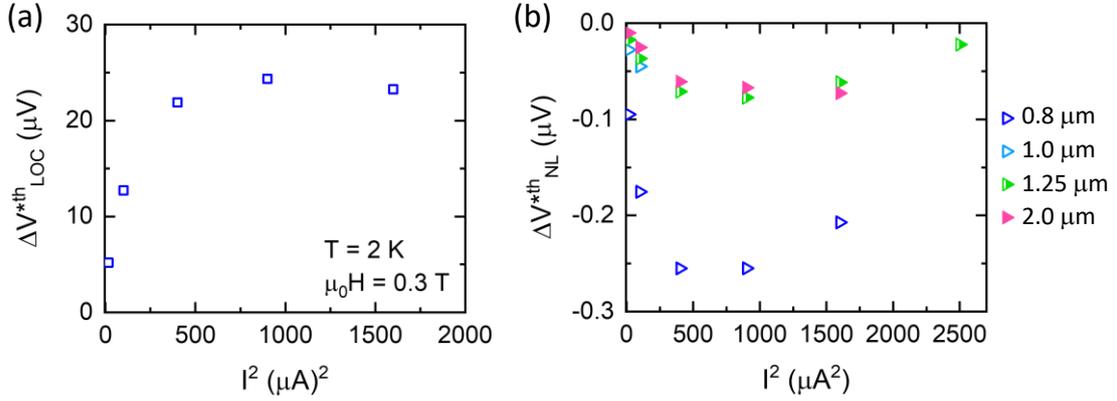

**FIG. S9**. Amplitude of the a) local SSE and b) non-local SSE as a function of the square of the current, at 2 K and 0.3 T.

### S3. Electrical excitation of magnons

The electrical excitation of magnons through the non-local voltage $V_{NL}^e$ have been verified for different injector-detector distances and no signal has been found. A representative measurement for device MST1 is shown in Fig. S5, in which we are not able to detect any $sin^2 \alpha$ modulation expected for the electrically driven magnons. The magnon population decreases as we approach zero temperature and thus the magnon accumulation and $V_{NL}^e$ vanishes [2–4]. The absence of an electrically driven signal can thus be explained by taking into account the range of temperatures involved (below ∼ 30 K) as well as the fact that $V_{NL}^e$ is linear with the applied current (while $V_{NL}^{th}$ is quadratic with the current, leading to larger values above the detection limit).

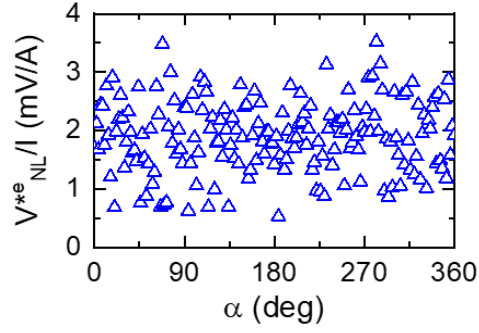

**FIG. S10.** Angular-dependent non-local signal detected for electrically injected magnon currents at d = 0.8 μm, $T$ = 2 K, I= 20 μA and $\mu_0 H$ = 0.3 T for device MST1.

### S4. Thermal excitation of magnons

Different regimes have been proposed for the propagation of magnons [5–7]. At short distances, $d \ll l_m^{th}$, the system is in the diffusive regime, and it decays as $1/d$. Then, for higher distances we enter in the exponential decay or relaxation regime. Finally, for $d \gg l_m^{th}$ the system enters the $1/d^2$ regime, where the signal reduction no longer depends on $l_m^{th}$. Figure S6 shows the different fittings according to these three regimes. From here, we conclude that the best fit is found for the exponential decay.

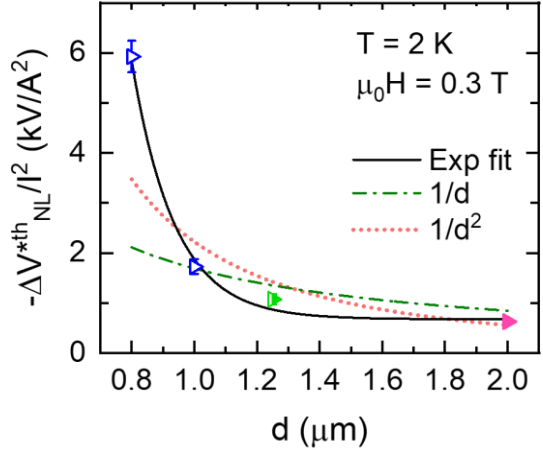

**FIG. S11.** Amplitude of the non-local SSE as a function of the injector-detector separation distance measured at $T$ = 2 K, $\mu_0 H$ = 0.3 T and $I$ = 4 μA. Black solid line is a fit to an exponential decay, green dashed line corresponds to $1/d$ fit and red dotted line to $1/d^2$ fit.

### S5. Magnetic characterization

The magnetic properties of the EuS thin film are determined by the vibrating sample magnetometry (VSM) and ferromagnetic resonance (FMR) techniques. Figure S7(a) shows the temperature dependence of the EuS magnetic moment ($m$) for different applied magnetic fields. EuS shows a clear ferromagnetic behavior with a broad transition to the paramagnetic state. From the temperature derivative of $m$, we found a Curie temperature $T_C \approx 19\ K$ for an applied magnetic field of 20 Oe [1,8]. Besides, an increase of the magnetic field applied shifts the transition to the paramagnetic state towards higher temperatures. The hysteresis loops recorded between 2 K and 30 K show very low coercive fields, around 3 mT at 2 K [see Fig. S6(b)], supporting the soft ferromagnetic behavior of the EuS films also observed in FDMR measurements [Fig. S1(a)]. For the probed magnetic field range, the magnetization increases with field, and we do not observe saturation.

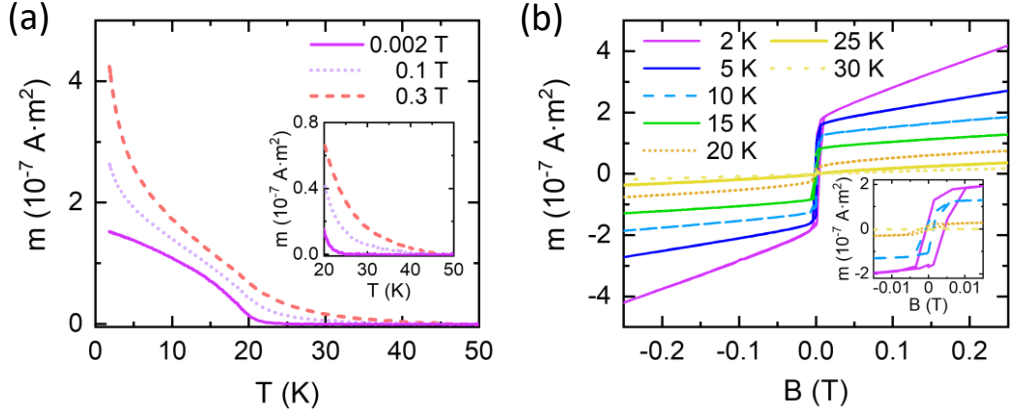

**FIG. S12.** (a) Magnetic moment of EuS thin film as a function of temperature for different applied magnetic fields $\mu_0 H$ = 2 mT, 0.1 T and 0.3 T. (b) Hysteresis loops measured at different temperatures from 2 K to 30 K with zoom to observe the small field regime, note that the serrated shape is due to the step size of 2.5 mT used during the measurement.

FMR measurements were performed at constant temperature by sweeping an external magnetic field at several fixed microwave frequencies in the range of 7-22 GHz. Each FMR spectrum has been analyzed by subtracting a background and fitting it to a Lorentzian curve [see Fig. S8(a)]. The resonance field ($H_{res}$) as a function of microwave frequency is shown in Fig. S8(b), which is fitted using the Kittel equation [solid lines in Fig. S8(b)]. According to Kittel formula, the precession frequency ($f$) is related to the material parameters by:

$$f = \frac{\gamma \mu_0}{2\pi} \sqrt{H_{res}(H_{res} + M_{eff})} \quad \text{(S2)}$$

where $\gamma$ is the gyromagnetic radio and $M_{eff}$ is the effective magnetization [9,10]. Due to a small magnetic anisotropy of the EuS we can assume that the saturation of magnetization $M_{sat} = M_{eff}$. As depicted in Fig. S8(c), the temperature behavior of $M_{eff}$ shows the same trend as the magnetic moment [see Fig. S7(a)], disappearing around $T_C$. At $T = 2$ K we found a saturation magnetization of 1163 kA/m, which is close to the reported bulk value of 1240 kA/m [11]. The FMR linewidth $\Delta H$ dependence on the excitation frequency is related to the Gilbert damping ($\alpha_G$), and it is described by the formula:

$$\Delta H = \Delta H_0 + \frac{4\pi \alpha_G}{\gamma} f \quad \text{(S3)}$$

where $\Delta H_0$ is termed the inhomogeneous linewidth [9]. From the linear fittings (solid lines in Fig. S8(d), we extract $\alpha_G = (40 \pm 4) \times 10^{-3}$ at 2 K. The values at other temperatures are plotted in Fig. S8(e), which is at least one order of magnitude higher than the damping of YIG.

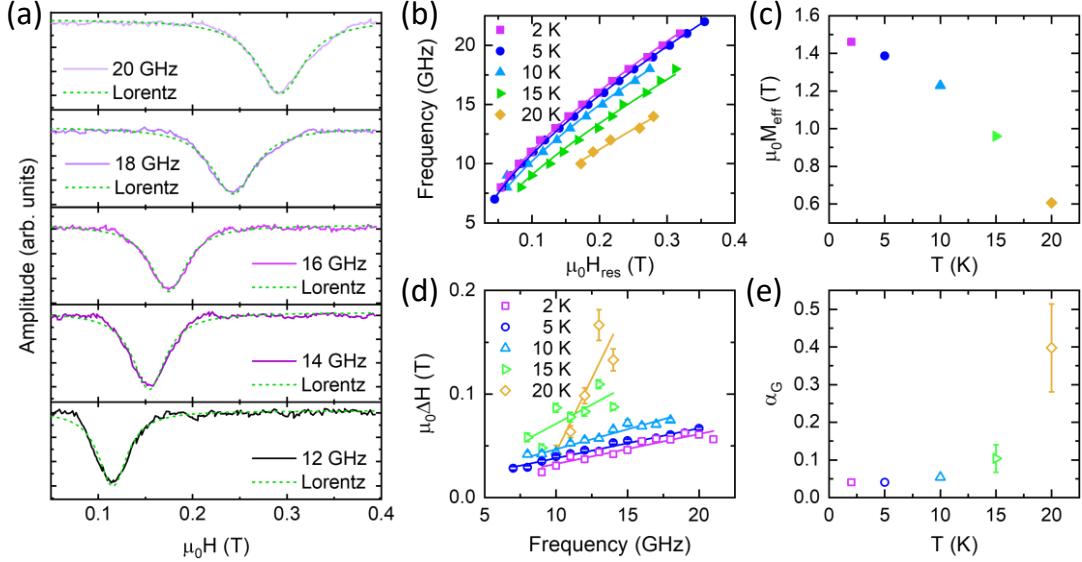

**FIG. S13.** (a) FMR spectras as a function of external magnetic field at 2 K for different frequencies. Dashed lines represent the fittings to a Lorentzian curve. (b) Frequency as a function of the resonant field at different temperatures. Solid lines correspond to the fit of experimental data to the Kittel formula (Eq. S2). (c) Effective saturation magnetization extracted from the fitting in panel (b). (d) FMR linewidth (ΔH) as a function of frequency. Solid lines correspond to a linear fit (see Eq. S3) of the experimental data. (e) Gilbert damping of the EuS sample as a function of temperature, extracted from the slope of panel (d).

## SUPPLEMENTAL REFERENCES